\documentclass[11pt]{article}
\usepackage[a4paper]{geometry}
\usepackage{amsfonts, amsmath, amssymb, amsthm, graphicx, caption, authblk, multirow, makecell, framed, float, xcolor, enumitem, tikz, hyperref, mathtools}
\theoremstyle{definition}

\theoremstyle{remark}

\newcommand*{\mybox}[1]{%
  \framebox{\raisebox{0cm}[0.5\baselineskip][0.05\baselineskip]{%
    \hbox to 0.10cm {\hss#1\hss}}}\hspace{0.05cm}}

\begin{document}
\title{Card-Based Overwriting Protocol for Equality Function and Applications}
\author[1]{Suthee Ruangwises\thanks{\texttt{ruangwises@uec.ac.jp}}}
\author[1]{Tomoki Ono\thanks{\texttt{onotom@uec.ac.jp}}}
\author[1,2]{Yoshiki Abe\thanks{\texttt{yoshiki@uec.ac.jp}}}
\author[1]{Kyosuke Hatsugai\thanks{\texttt{hatsugai@uec.ac.jp}}}
\author[1]{Mitsugu Iwamoto\thanks{\texttt{mitsugu@uec.ac.jp}}}
\affil[1]{The University of Electro-Communications, Tokyo, Japan}
\affil[2]{National Institute of Advanced Industrial Science and Technology, Tokyo, Japan}
\date{}
\maketitle

\begin{abstract}
Research in the area of secure multi-party computation with an unconventional method of using a physical deck of playing cards began in 1989 when den Boer proposed a protocol to compute the logical AND function using five cards. Since then, the area has gained interest from many researchers and several card-based protocols to compute various functions have been developed. In this paper, we propose a card-based protocol called the \textit{overwriting protocol} that can securely compute the $k$-candidate $n$-variable \textit{equality function} $f: \{0,1,\ldots ,k-1\}^n \rightarrow \{0,1\}$. We also apply the technique used in this protocol to compute other similar functions.

\textbf{Keywords:} card-based cryptography, secure multi-party computation, equality function
\end{abstract}

\section{Introduction}
During a presidential election with $k$ candidates, a group of $n$ friends decide that they will talk about politics only if all of them support the same candidate. However, each person does not want to reveal his/her preference to the group (unless it is known that everyone supports the same candidate). They need a way to find out whether their preferences all coincide without leaking information about any individual's preference (not even probabilistic information).

Theoretically, this is equivalent to each $i$-th person having an integer $a_i \in \{0,1,\ldots ,k-1\}$, indicating the candidate he/she prefers. Define an \textit{equality function} $E(a_1,a_2,\ldots ,a_n) \coloneqq 1$ if $a_1=a_2=\cdots =a_n$, and $E(a_1,a_2,\ldots ,a_n) \coloneqq 0$ otherwise. Our goal is to develop a protocol that can compute the value of $E(a_1,a_2,\ldots ,a_n)$ without leaking any other information.

This situation is an example of secure multi-party computation, one of the most actively studied areas in cryptography, which studies how multiple parties can compare their private information without revealing it. In contrast to digital protocols, unconventional protocols using physical objects such as coins \cite{coin} and combination locks \cite{diallock} have also been studied, but the most used object is a deck of playing cards. Thus, this research area is often called card-based cryptography.

Card-based protocols have benefits that they do not require computers, and also allow external observers to verify that all parties truthfully execute the protocol (which is often a challenging task for digital protocols). They also have a great didactic value and thus can be used to teach the concept of secure multi-party computation to non-experts.

\subsection{Related Work}
Research in card-based cryptography began in 1989 when den Boer \cite{5card} proposed a protocol called the \textit{five-card trick} to compute the logical AND function of two players' bits using five cards. After that, other AND function protocols \cite{abe,crepeau,koch21,koch15,mizuki12,mizuki09,niemi,ruangwises,stiglic} were also developed. These subsequent protocols either reduced the number of required cards or shuffles, or improved properties of the protocol involving output format, type of shuffles, etc.

Apart from the AND function protocols, protocols to compute other functions have been developed as well, including XOR function protocols \cite{crepeau,mizuki09,mizuki06}, \textit{majority function} protocols \cite{nishida13,toyoda} (deciding whether there are more 1s than 0s in the input bits), a copy protocol \cite{mizuki09} (duplicating the input), and a voting protocol \cite{mizuki13} (adding input bits and storing the sum in binary representation). Nishida et al. \cite{nishida15} proved that any $n$-variable Boolean function can be computed using $2n+6$ cards, and any such function that is symmetric can be computed using $2n+2$ cards. In fact, many classes of such symmetric functions can be optimally computed using $2n$ cards \cite{ruangwises3}.

While almost all of the existing card-based protocols were designed to compute Boolean functions, a few results also focused on computing functions in $\mathbb{Z}/k\mathbb{Z}$ for $k>2$, such as \cite{z6,triangle,polygon}.

\subsubsection{Equality Function Protocols}
Several previous results have been focused on computing the equality function. For the Boolean equality function (where $k=2$), a protocol for a special case $n=3$ has been independently proposed by~Heather et al. \cite{6card0} and by Shinagawa and Mizuki \cite{6card}. Later, Ruangwises and Itoh \cite{ruangwises2} developed two protocols to compute the Boolean equality function for any $n$. Their work can also be generalized to compute the $k$-candidate $n$-variable equality function $f: \{0,1,\ldots ,k-1\}^n \rightarrow \{0,1\}$ for any $k$ and $n$ by converting each input into binary representation.

\subsection{Our Contribution}
In this paper, we propose a new card-based protocol called the \textit{overwriting protocol}, which can securely compute the $k$-candidate $n$-variable equality function $f: \{0,1,\ldots ,k-1\}^n \rightarrow \{0,1\}$ using $kn$ cards and $n$ shuffles.

Comparing to the existing protocol of Ruangwises and Itoh \cite{ruangwises2} which uses $2 \lceil \lg k \rceil n$ cards and $\lceil \lg k \rceil n-1$ shuffles, our protocol uses fewer cards when $k=3$ or 5 (and uses equal number of cards when $k=2$, 4, or 6), and also uses fewer shuffles for every $k \geq 3$. See Table~\ref{table1} for comparison. In addition, our protocol is simpler and more intuitive, as it does not require each player to convert an integer into binary representation.

We also apply the overwriting technique used in our protocol to compute two other functions: the \textit{set function} and the \textit{set size function} (where the definitions are given in Section~\ref{app}). See Table~\ref{table2} for comparison.

\begin{table}
	\centering
	\caption{Number of required cards and shuffles for each equaility function protocol} \label{table1}
	\begin{tabular}{|c|c|c|c|}
		\hline
		\textbf{Protocol} & \textbf{Equality Function} & \textbf{\#Cards} & \textbf{\#Shuffles} \\ \hline
		Heather et al. \cite{6card0} & \multirow{2}{*}{$E:\{0,1\}^3 \rightarrow \{0,1\}$} & 6 & 1 \\ \cline{1-1} \cline{3-4}
		Shinagawa-Mizuki \cite{6card} & & 6 & 1 \\ \hline
		Ruangwises-Itoh \cite{ruangwises2} & $E:\{0,1\}^n \rightarrow \{0,1\}$ & $2n$ & $n-1$ \\ \hline
		Ruangwises-Itoh \cite{ruangwises2} & \multirow{2}{*}{$E:\{0,1,\ldots ,k-1\}^n \rightarrow \{0,1\}$} & $2 \lceil \lg k \rceil n$ & $\lceil \lg k \rceil n-1$ \\ \cline{1-1} \cline{3-4}
		\textbf{Ours (\S\ref{main})} & & $kn$ & $n$ \\ \hline
	\end{tabular}
	\medskip
\end{table}

\begin{table}
	\centering
	\caption{Number of required cards and shuffles for each of our three protocols} \label{table2}
	\begin{tabular}{|c|c|c|c|}
		\hline
		\textbf{Protocol} & \textbf{Function} & \textbf{\#Cards} & \textbf{\#Shuffles} \\ \hline
		\textbf{Equality (\S\ref{main})} & $E:\{0,1,\ldots ,k-1\}^n \rightarrow \{0,1\}$ & $kn$ & $n$ \\ \hline
		\textbf{Set Size (\S\ref{setsize})} & $SS:\{0,1,\ldots ,k-1\}^n \rightarrow \{0,1,\ldots ,k\}$ & $kn$ & $n$ \\ \hline
		\textbf{Set (\S\ref{set})} & $S:\{0,1,\ldots ,k-1\}^n \rightarrow \mathcal{P}(\{0,1,\ldots ,k-1\})$ & $k(n+1)$ & $n$ \\ \hline
	\end{tabular}
	\medskip
\end{table}

\section{Preliminaries}
\subsection{Equality Function}
Define an equality function $E:\{0,1,\ldots ,k-1\}^n \rightarrow \{0,1\}$ as
$$E(a_1,a_2,\ldots ,a_n) \coloneqq
\begin{cases}
	1, & \text{if $a_1=a_2=\cdots =a_n$;} \\
	0, & \text{otherwise.}
\end{cases}$$

\subsection{Cards}
Each card used in our protocol has either a $\clubsuit$ or a $\heartsuit$ on the front side. All cards have indistinguishable back sides.

An integer $i \in \{0,1,\ldots ,k-1\}$ is encoded by a sequence of $k$ consecutive cards, all of them being \mybox{$\heartsuit$}s except the ($i+1$)-th leftmost card being a \mybox{$\clubsuit$}. Such a sequence is called $E_k(i)$, e.g. $E_4(1)$ is \mybox{$\heartsuit$}\mybox{$\clubsuit$}\mybox{$\heartsuit$}\mybox{$\heartsuit$}.

\subsection{Pile-Scramble Shuffle}
A \textit{pile-scramble shuffle} \cite{scramble} rearranges the columns of an $n \times k$ matrix of cards by a uniformly random permutation unknown to all parties, i.e. moves each Column $i$ to Column $p_i$ for a uniformly random permutation $p=(p_1,p_2,\ldots ,p_k)$ of $(1,2,\ldots ,k)$. See Fig.~\ref{fig3}.

It can be implemented in real world by putting all cards in each column into an envelope, and scrambling all envelopes together completely randomly on a table.

\begin{figure}[H]
\centering
\begin{tikzpicture}
\node at (0,0) {\mybox{?}};
\node at (0.5,0) {\mybox{?}};
\node at (1,0) {\mybox{?}};
\node at (1.5,0) {\mybox{?}};
\node at (2,0) {\mybox{?}};
\node at (2.5,0) {\mybox{?}};

\node at (0,0.6) {\mybox{?}};
\node at (0.5,0.6) {\mybox{?}};
\node at (1,0.6) {\mybox{?}};
\node at (1.5,0.6) {\mybox{?}};
\node at (2,0.6) {\mybox{?}};
\node at (2.5,0.6) {\mybox{?}};

\node at (0,1.2) {\mybox{?}};
\node at (0.5,1.2) {\mybox{?}};
\node at (1,1.2) {\mybox{?}};
\node at (1.5,1.2) {\mybox{?}};
\node at (2,1.2) {\mybox{?}};
\node at (2.5,1.2) {\mybox{?}};

\node at (0,1.8) {\mybox{?}};
\node at (0.5,1.8) {\mybox{?}};
\node at (1,1.8) {\mybox{?}};
\node at (1.5,1.8) {\mybox{?}};
\node at (2,1.8) {\mybox{?}};
\node at (2.5,1.8) {\mybox{?}};

\node at (0,2.4) {\mybox{?}};
\node at (0.5,2.4) {\mybox{?}};
\node at (1,2.4) {\mybox{?}};
\node at (1.5,2.4) {\mybox{?}};
\node at (2,2.4) {\mybox{?}};
\node at (2.5,2.4) {\mybox{?}};

\node at (-0.4,0) {5};
\node at (-0.4,0.6) {4};
\node at (-0.4,1.2) {3};
\node at (-0.4,1.8) {2};
\node at (-0.4,2.4) {1};

\node at (0,2.9) {1};
\node at (0.5,2.9) {2};
\node at (1,2.9) {3};
\node at (1.5,2.9) {4};
\node at (2,2.9) {5};
\node at (2.5,2.9) {6};

\node at (3.4,1.2) {\LARGE{$\Rightarrow$}};
\end{tikzpicture}
\begin{tikzpicture}
\node at (0,0) {\mybox{?}};
\node at (0.5,0) {\mybox{?}};
\node at (1,0) {\mybox{?}};
\node at (1.5,0) {\mybox{?}};
\node at (2,0) {\mybox{?}};
\node at (2.5,0) {\mybox{?}};

\node at (0,0.6) {\mybox{?}};
\node at (0.5,0.6) {\mybox{?}};
\node at (1,0.6) {\mybox{?}};
\node at (1.5,0.6) {\mybox{?}};
\node at (2,0.6) {\mybox{?}};
\node at (2.5,0.6) {\mybox{?}};

\node at (0,1.2) {\mybox{?}};
\node at (0.5,1.2) {\mybox{?}};
\node at (1,1.2) {\mybox{?}};
\node at (1.5,1.2) {\mybox{?}};
\node at (2,1.2) {\mybox{?}};
\node at (2.5,1.2) {\mybox{?}};

\node at (0,1.8) {\mybox{?}};
\node at (0.5,1.8) {\mybox{?}};
\node at (1,1.8) {\mybox{?}};
\node at (1.5,1.8) {\mybox{?}};
\node at (2,1.8) {\mybox{?}};
\node at (2.5,1.8) {\mybox{?}};

\node at (0,2.4) {\mybox{?}};
\node at (0.5,2.4) {\mybox{?}};
\node at (1,2.4) {\mybox{?}};
\node at (1.5,2.4) {\mybox{?}};
\node at (2,2.4) {\mybox{?}};
\node at (2.5,2.4) {\mybox{?}};

\node at (-0.4,0) {5};
\node at (-0.4,0.6) {4};
\node at (-0.4,1.2) {3};
\node at (-0.4,1.8) {2};
\node at (-0.4,2.4) {1};

\node at (0,2.9) {3};
\node at (0.5,2.9) {6};
\node at (1,2.9) {4};
\node at (1.5,2.9) {1};
\node at (2,2.9) {5};
\node at (2.5,2.9) {2};
\end{tikzpicture}
\caption{An example of a pile-scramble shuffle on a $5 \times 6$ matrix}
\label{fig3}
\end{figure}

\subsection{Pile-Shifting Shuffle}
A \textit{pile-shifting shuffle} \cite{polygon} rearranges the columns of an $n \times k$ matrix of cards by a random cyclic shift unknown to all parties, i.e. moves each Column $i$ to Column $i+r$ for a uniformly random $r \in \{0,1,\ldots ,k-1\}$ (where Column $j$ means Column $j-k$ for $j \geq k$). See Fig.~\ref{fig4}.

It can be implemented in real world by putting the cards in each column into an envelope, and taking turns to apply \textit{Hindu cuts} (taking several envelopes from the bottom and putting them on the top), to the pile of envelopes \cite{hindu}.

\begin{figure}[H]
\centering
\begin{tikzpicture}
\node at (0,0) {\mybox{?}};
\node at (0.5,0) {\mybox{?}};
\node at (1,0) {\mybox{?}};
\node at (1.5,0) {\mybox{?}};
\node at (2,0) {\mybox{?}};
\node at (2.5,0) {\mybox{?}};

\node at (0,0.6) {\mybox{?}};
\node at (0.5,0.6) {\mybox{?}};
\node at (1,0.6) {\mybox{?}};
\node at (1.5,0.6) {\mybox{?}};
\node at (2,0.6) {\mybox{?}};
\node at (2.5,0.6) {\mybox{?}};

\node at (0,1.2) {\mybox{?}};
\node at (0.5,1.2) {\mybox{?}};
\node at (1,1.2) {\mybox{?}};
\node at (1.5,1.2) {\mybox{?}};
\node at (2,1.2) {\mybox{?}};
\node at (2.5,1.2) {\mybox{?}};

\node at (0,1.8) {\mybox{?}};
\node at (0.5,1.8) {\mybox{?}};
\node at (1,1.8) {\mybox{?}};
\node at (1.5,1.8) {\mybox{?}};
\node at (2,1.8) {\mybox{?}};
\node at (2.5,1.8) {\mybox{?}};

\node at (0,2.4) {\mybox{?}};
\node at (0.5,2.4) {\mybox{?}};
\node at (1,2.4) {\mybox{?}};
\node at (1.5,2.4) {\mybox{?}};
\node at (2,2.4) {\mybox{?}};
\node at (2.5,2.4) {\mybox{?}};

\node at (-0.4,0) {5};
\node at (-0.4,0.6) {4};
\node at (-0.4,1.2) {3};
\node at (-0.4,1.8) {2};
\node at (-0.4,2.4) {1};

\node at (0,2.9) {1};
\node at (0.5,2.9) {2};
\node at (1,2.9) {3};
\node at (1.5,2.9) {4};
\node at (2,2.9) {5};
\node at (2.5,2.9) {6};

\node at (3.4,1.2) {\LARGE{$\Rightarrow$}};
\end{tikzpicture}
\begin{tikzpicture}
\node at (0,0) {\mybox{?}};
\node at (0.5,0) {\mybox{?}};
\node at (1,0) {\mybox{?}};
\node at (1.5,0) {\mybox{?}};
\node at (2,0) {\mybox{?}};
\node at (2.5,0) {\mybox{?}};

\node at (0,0.6) {\mybox{?}};
\node at (0.5,0.6) {\mybox{?}};
\node at (1,0.6) {\mybox{?}};
\node at (1.5,0.6) {\mybox{?}};
\node at (2,0.6) {\mybox{?}};
\node at (2.5,0.6) {\mybox{?}};

\node at (0,1.2) {\mybox{?}};
\node at (0.5,1.2) {\mybox{?}};
\node at (1,1.2) {\mybox{?}};
\node at (1.5,1.2) {\mybox{?}};
\node at (2,1.2) {\mybox{?}};
\node at (2.5,1.2) {\mybox{?}};

\node at (0,1.8) {\mybox{?}};
\node at (0.5,1.8) {\mybox{?}};
\node at (1,1.8) {\mybox{?}};
\node at (1.5,1.8) {\mybox{?}};
\node at (2,1.8) {\mybox{?}};
\node at (2.5,1.8) {\mybox{?}};

\node at (0,2.4) {\mybox{?}};
\node at (0.5,2.4) {\mybox{?}};
\node at (1,2.4) {\mybox{?}};
\node at (1.5,2.4) {\mybox{?}};
\node at (2,2.4) {\mybox{?}};
\node at (2.5,2.4) {\mybox{?}};

\node at (-0.4,0) {5};
\node at (-0.4,0.6) {4};
\node at (-0.4,1.2) {3};
\node at (-0.4,1.8) {2};
\node at (-0.4,2.4) {1};

\node at (0,2.9) {5};
\node at (0.5,2.9) {6};
\node at (1,2.9) {1};
\node at (1.5,2.9) {2};
\node at (2,2.9) {3};
\node at (2.5,2.9) {4};
\end{tikzpicture}
\caption{An example of a pile-shifting shuffle on a $5 \times 6$ matrix}
\label{fig4}
\end{figure}

\section{Protocol for Equality Function} \label{main}
Each player is given $k$ cards: one \mybox{$\clubsuit$} and $k-1$ \mybox{$\heartsuit$}s. Each $i$-th player secretly arranges the given cards as a face-down sequence $E_k(a_i)$, encoding the value of $a_i$. Then, all players together perform the following steps.

\begin{figure}
\centering
\begin{tikzpicture}
\node at (0,0) {\mybox{$\heartsuit$}};
\node at (0.5,0) {\mybox{$\heartsuit$}};
\node at (1,0) {\mybox{$\clubsuit$}};
\node at (1.5,0) {\mybox{$\heartsuit$}};
\node at (2,0) {\mybox{$\heartsuit$}};
\node at (2.5,0) {\mybox{$\heartsuit$}};

\node at (0,0.6) {\mybox{$\clubsuit$}};
\node at (0.5,0.6) {\mybox{$\heartsuit$}};
\node at (1,0.6) {\mybox{$\heartsuit$}};
\node at (1.5,0.6) {\mybox{$\heartsuit$}};
\node at (2,0.6) {\mybox{$\heartsuit$}};
\node at (2.5,0.6) {\mybox{$\heartsuit$}};

\node at (0,1.2) {\mybox{$\heartsuit$}};
\node at (0.5,1.2) {\mybox{$\heartsuit$}};
\node at (1,1.2) {\mybox{$\clubsuit$}};
\node at (1.5,1.2) {\mybox{$\heartsuit$}};
\node at (2,1.2) {\mybox{$\heartsuit$}};
\node at (2.5,1.2) {\mybox{$\heartsuit$}};

\node at (0,1.8) {\mybox{$\heartsuit$}};
\node at (0.5,1.8) {\mybox{$\heartsuit$}};
\node at (1,1.8) {\mybox{$\heartsuit$}};
\node at (1.5,1.8) {\mybox{$\clubsuit$}};
\node at (2,1.8) {\mybox{$\heartsuit$}};
\node at (2.5,1.8) {\mybox{$\heartsuit$}};

\node at (0,2.4) {\mybox{$\heartsuit$}};
\node at (0.5,2.4) {\mybox{$\heartsuit$}};
\node at (1,2.4) {\mybox{$\clubsuit$}};
\node at (1.5,2.4) {\mybox{$\heartsuit$}};
\node at (2,2.4) {\mybox{$\heartsuit$}};
\node at (2.5,2.4) {\mybox{$\heartsuit$}};

\node at (-0.4,0) {5};
\node at (-0.4,0.6) {4};
\node at (-0.4,1.2) {3};
\node at (-0.4,1.8) {2};
\node at (-0.4,2.4) {1};

\node at (0,2.9) {1};
\node at (0.5,2.9) {2};
\node at (1,2.9) {3};
\node at (1.5,2.9) {4};
\node at (2,2.9) {5};
\node at (2.5,2.9) {6};
\end{tikzpicture}
\caption{An example of a matrix $M$ constructed in Step 1, where $k=6$, $n=5$, $a_1=a_3=a_5=2$, $a_2=3$, and $a_4=0$ (with all cards actually being face-down)}
\label{fig5}
\end{figure}

\begin{enumerate}
	\item Construct an $n \times k$ matrix $M$ by putting a sequence $E_k(a_i)$ in Row $i$ of $M$ for each $i=1,2,\ldots ,n$. See an example in Fig.~\ref{fig5}. Let $M(x,y)$ denotes a card at Row $x$ and Column $y$ of $M$ (the indices start at 1, not 0).
	\item Perform the following steps for $i=2,3,\ldots ,n$.
	\begin{enumerate}
		\item Apply the pile-scramble shuffle to $M$.
		\item Turn over all cards in Row $i$ of $M$.
		\item For each $j=1,2,\ldots ,k$, if $M(i,j)$ is a \mybox{$\heartsuit$}, swap $M(i,j)$ and $M(1,j)$.
		\item Turn over all face-up cards.
	\end{enumerate}
	\item Apply the pile-scramble shuffle to $M$.
	\item Turn over all cards in Row $1$ of $M$. If the cards consist of one \mybox{$\clubsuit$} and $k-1$~\mybox{$\heartsuit$}s, output 1; otherwise, output 0.
\end{enumerate}

\section{Proofs of Correctness and Security}
\subsection{Proof of Correctness} \label{cor}
Observe that in Step 2(c), we ``overwrite'' a card $M(1,j)$ in Row 1, replacing it with a \mybox{$\heartsuit$}. Therefore, the total number of \mybox{$\heartsuit$}s in Row 1 never decreases throughout the protocol. Hence, there are only two possibilities of the cards in Row 1 at the end: one \mybox{$\clubsuit$} and $k-1$ \mybox{$\heartsuit$}s, or all $k$ \mybox{$\heartsuit$}s.

The only case where Row 1 consists of one \mybox{$\clubsuit$} and $k-1$ \mybox{$\heartsuit$}s at the end occurs if and only if we never replace a \mybox{$\clubsuit$} in Row 1 with a \mybox{$\heartsuit$} in any step throughout the protocol. This condition occurs when a \mybox{$\clubsuit$} in Row $i$ lies in the same column as a \mybox{$\clubsuit$} in Row 1 for every $i=2,3,\ldots ,n$, i.e. $a_1=a_2=\ldots =a_n$, which is exactly the condition where the equality function outputs 1. Hence, our protocol is always correct.

\subsection{Proof of Security} \label{sec}
It is sufficient to show that any step that reveals face-up cards does not leak any information about the inputs.

Consider Step 2(b). The cards in Row $i$ have never been swapped with cards from other row before, so they consist of one \mybox{$\clubsuit$} and $k-1$ \mybox{$\heartsuit$}s. Moreover, due to the pile-scramble shuffle in Step 2(a), the only \mybox{$\clubsuit$} has an equal probability of $1/k$ to be in each column no matter what $a_i$ is. Therefore, this step does not leak information about the inputs.

Consider Step 4. As proved in Section~\ref{cor}, Row 1 either consists of all $k$~\mybox{$\heartsuit$}s, or one \mybox{$\clubsuit$} and $k-1$ \mybox{$\heartsuit$}s. In the latter case, due to the pile-scramble shuffle in Step 3, the only \mybox{$\clubsuit$} has an equal probability of $1/k$ to be in each column. Therefore, this step only reveals the value of $E(a_1,a_2,\ldots ,a_n)$ without leaking other information about the inputs.

\section{Applications to Set and Set Size Functions} \label{app}
Suppose during a job recruitment, a total of $k$ candidates have applied for the position. Each of the $n$ referees secretly selects a candidate he/she prefers. Candidate(s) that are selected by at least one referee will advance to the final round interview. The referees need a method to find out which candidate(s) will advance to the final round without leaking the preference of any individual referee.

This situation is similar to that of the equality function where each $i$-th referee has an integer $a_i \in \{0,1,\ldots ,k-1\}$ indicating the candidate he/she prefers, but we instead want to find a set of integers that are equal to at least one input value. Define a \textit{set function} $S:\{0,1,\ldots ,k-1\}^n \rightarrow \mathcal{P}(\{0,1,\ldots ,k-1\})$ as
$$S(a_1,a_2,\ldots ,a_n) \coloneqq \{j | \exists i, a_i=j\},$$
where $\mathcal{P}$ denotes the power set. For example, $S(3,2,3,0,5,0)=\{0,2,3,5\}$. Our goal is to develop a protocol to securely compute the set function between $n$ players.

A closely related problem is when we want to find only the number of candidates that will advance to the final round (e.g. for the purpose of venue preparation), but not \textit{which} candidate(s). In this case, define a \textit{set size function} $SS:\{0,1,\ldots ,k-1\}^n \rightarrow \{0,1,\ldots ,k\}$ as
\begin{align*}
SS(a_1,a_2,\ldots ,a_n) \coloneqq &|S(a_1,a_2,\ldots ,a_n)|\\
= &|\{j | \exists i, a_i=j\}|.
\end{align*}
For example, $SS(3,2,3,0,5,0)=|\{0,2,3,5\}|=4$. Note that $SS(a_1,a_2,\ldots ,a_n)=1$ if and only if $E(a_1,a_2,\ldots ,a_n) = 1$. We also want to develop a protocol for the set size function between $n$ players.

It turns out that a protocol for the set size function is very similar to the one for the equality function, so we will introduce it first. Then, we will introduce a protocol for the set function.

\subsection{Protocol for Set Size Function} \label{setsize}
Luckily, we can slightly modify the equality function protocol in Section~\ref{main} to compute the set size function. The setup phase is the same, with each $i$-th player having a sequence $E_k(a_i)$. In the modified procotol, all players together perform the following steps.

\begin{enumerate}
	\item Construct an $n \times k$ matrix $M$ by putting a sequence $E_k(a_i)$ in Row $i$ of $M$ for each $i=1,2,\ldots ,n$.
	\item Perform the following steps for $i=2,3,\ldots ,n$.
	\begin{enumerate}
		\item Apply the pile-scramble shuffle to $M$.
		\item Turn over all cards in Row $i$ of $M$.
		\item For each $j=1,2,\ldots ,k$, if $M(i,j)$ is a \mybox{$\clubsuit$}, swap $M(i,j)$ and $M(1,j)$.
		\item Turn over all face-up cards.
	\end{enumerate}
	\item Apply the pile-scramble shuffle to $M$.
	\item Turn over all cards in Row $1$ of $M$. Output the total number of \mybox{$\clubsuit$}s in Row~1.
\end{enumerate}

The only differences from the equality function protocol are that in Step~2(c), we replace a card $M(1,j)$ by a \mybox{$\clubsuit$} instead of a \mybox{$\heartsuit$}, and that in Step 4, we output the total number of \mybox{$\clubsuit$}s in Row 1.

\subsubsection{Proof of Correctness} \label{cor2}
Observe that in Step 2(c), we ``overwrite'' a card $M(1,j)$ in Row 1, replacing it with a \mybox{$\clubsuit$}. Therefore, once a card in Row 1 becomes a \mybox{$\clubsuit$}, it will always stay a \mybox{$\clubsuit$} thoughout the protocol.

Consider a card $M(1,j)$ at the beginning of the protocol (we consider this exact card no matter which column it later moves to). If it is a \mybox{$\clubsuit$} at the beginning, i.e. $a_1=j-1$, it will stay a \mybox{$\clubsuit$} at the end. On the other hand, if it is a~\mybox{$\heartsuit$} at the beginnning, it will become a \mybox{$\clubsuit$} at the end if and only if it has been replaced by a \mybox{$\clubsuit$} from Row $i \geq 2$ at some step during the protocol, which occurs when $a_i=j-1$.

Therefore, we can conclude that a card $M(1,j)$ at the beginning becomes a~\mybox{$\clubsuit$} at the end if and only if there is an index $i \in \{1,2,\ldots ,n\}$ such that $a_i=j-1$. This means the number of \mybox{$\clubsuit$}s in Row 1 at the end is equal to the number of such indices $j$, which is exactly the output of the set size function. Hence, our protocol is always correct.

\subsubsection{Proof of Security}
It is sufficient to show that any step that reveals face-up cards does not leak any information about the inputs.

Consider Step 2(b). The cards in Row $i$ have never been swapped with cards from other row before, so they consist of one \mybox{$\clubsuit$} and $k-1$ \mybox{$\heartsuit$}s. Moreover, due to the pile-scramble shuffle in Step 2(a), the only \mybox{$\clubsuit$} has an equal probability of $1/k$ to be in each column no matter what $a_i$ is. Therefore, this step does not leak information about the inputs.

Consider Step 4. Suppose Row 1 consists of $\ell$ \mybox{$\clubsuit$}s and $k-\ell$ \mybox{$\heartsuit$}s. Due to the pile-scramble shuffle in Step 3, the $\ell$ \mybox{$\clubsuit$}s have an equal probability to be in each of the $\binom{k}{\ell}$ combinations of columns. Therefore, this step only reveals the value $\ell = SS(a_1,a_2,\ldots ,a_n)$ without leaking other information about the inputs.

\subsection{Protocol for Set Function} \label{set}
We modify the set size function protocol in Section~\ref{setsize} to compute the set function. The setup phase is the same, with each $i$-th player having a sequence $E_k(a_i)$. All players together perform the following steps.

\begin{enumerate}
	\item Construct an $(n+1) \times k$ matrix $M$ by putting a sequence $E_k(a_i)$ in Row $i$ of $M$ for each $i=1,2,\ldots ,n$ and putting a sequence $E_k(0)$ in Row $n+1$.
	\item Perform the following steps for $i=2,3,\ldots ,n$.
	\begin{enumerate}
		\item Apply the pile-shifting shuffle to $M$.
		\item Turn over all cards in Row $i$ of $M$.
		\item For each $j=1,2,\ldots ,k$, if $M(i,j)$ is a \mybox{$\clubsuit$}, swap $M(i,j)$ and $M(1,j)$.
		\item Turn over all face-up cards.
	\end{enumerate}
	\item Apply the pile-shifting shuffle to $M$.
	\item Turn over all cards in Row $n+1$ of $M$. Shift the columns of $M$ cyclically such that the only \mybox{$\clubsuit$} in Row $n+1$ moves to Column 1.
	\item Turn over all cards in Row $1$ of $M$. Output the set $\{j|M(1,j+1)\text{ is a \mybox{$\clubsuit$}}\}$.
\end{enumerate}

Note that in this protocol, we apply the pile-shifting shuffle instead of the pile-scramble shuffle in Steps 2(a) and 3. The purpose of Row $n+1$ and Step 4 is to revert the matrix $M$ to its original position before any pile-shifting shuffle.

\subsubsection{Proof of Correctness}
By the same reasons as the proof of correctness in Section~\ref{cor2}, a card $M(1,j)$ at the beginning becomes a \mybox{$\clubsuit$} at the end if and only if there is an index $i \in \{1,2,\ldots ,n\}$ such that $a_i=j-1$.

In this protocol, we only apply pile-shifting shuffles without pile-scramble shuffles, so the columns of $M$ only move cyclically throughout the protocol. Thus, in Step 4 where the \mybox{$\clubsuit$} in Row $n+1$ returns to its original position in Column 1, the columns of $M$ must also return to their original order at the beginning of the protocol. This means the set outputted in Step 5 is exactly the set $\{j | \exists i, a_i=j\}$. Hence, our protocol is always correct.

\subsubsection{Proof of Security}
It is sufficient to show that any step that reveals face-up cards does not leak any information about the inputs.

Consider Step 2(b). The cards in Row $i$ have never been swapped with cards from other row before, so they consist of one \mybox{$\clubsuit$} and $k-1$ \mybox{$\heartsuit$}s. Moreover, due to the pile-shifting shuffle in Step 2(a), the only \mybox{$\clubsuit$} has an equal probability of $1/k$ to be in each column no matter what $a_i$ is. Therefore, this step does not leak information about the inputs.

Consider Step 4. The cards in Row $n+1$ have never been swapped with cards from other row before, so they consist of one \mybox{$\clubsuit$} and $k-1$ \mybox{$\heartsuit$}s. Moreover, due to the pile-shifting shuffle in Step 3, the only \mybox{$\clubsuit$} has an equal probability of $1/k$ to be in each column. Therefore, this step does not leak information about the inputs.

Consider Step 5. The positions of all \mybox{$\clubsuit$}s in Row 1 only reveal the set $\{j | \exists i, a_i=j\}$, which is exactly the output of the set function. Therefore, this step does not leak information about the inputs.

\section{Future Work}
We developed three protocols to compute the equality, set, and set size functions. Possible future work includes reducing the number of required cards and shuffles in these protocols, or proving the lower bounds of them. Also, as all results so far have been focusing on using two types of cards, it is an interesting question that whether we can reduce the number of cards if three or more types of cards are used. Other challenging work includes developing protocols to compute other functions in $\mathbb{Z}/k\mathbb{Z}$ for $k>2$.

\end{document}